\documentclass[aps,prb,twocolumn,superscriptaddress,showpacs]{revtex4-1}
\usepackage{graphicx}
\usepackage{dcolumn}
\usepackage{bm}
\usepackage{amsfonts}
\usepackage{amsmath}
\usepackage{ulem}
\usepackage{color}

\begin{document}

\title{Tuning Excess Noise by Aharonov-Bohm Interferometry\footnote{{\it Article dedicated to Peter H{\"a}nggi on occasion of his sixtieth birthday}}}

\author{Fabrizio Dolcini}
\email{fabrizio.dolcini@polito.it}
\affiliation{Dipartimento di Fisica del Politecnico di Torino, I-10129 Torino, Italy}
\author{Hermann Grabert}
\
\affiliation{Physikalisches Institut, Albert-Ludwigs-Universit\"at, 79104 Freiburg,
Germany}
\affiliation{Freiburg Institute for
Advanced Studies (FRIAS), Albert-Ludwigs-Universit\"at, 79104 Freiburg,
Germany}

\begin{abstract}
A voltage bias applied to a conductor induces a change of the current noise with respect to the equilibrium noise known as excess noise. We analyze the excess noise of the electronic current flowing through a mesoscopic Aharonov-Bohm ring threaded by a magnetic flux, coupled to a side gate, and contacted by two metallic electrodes. It is shown that the excess noise can be controlled both magnetically and electrostatically, demonstrating the full tunability of the system. At zero frequency, the ratio of the noise strength to the current (Fano factor) can thereby be minimized. Remarkably, at finite frequency, regions of negative excess noise emerge.
\end{abstract}


\maketitle


\section{Introduction}
\noindent  Electronic conduction is an important phenomenon arising in a broad variety of  systems in different fields, ranging from physics to engineering and chemistry. Conductors can be organic and inorganic materials, and there are bulk materials like metals, molecular systems such as polymers or DNA, as well as micro-fabricated devices. The most important property of a conductor is its current-voltage characteristics, i.e. the dependence of the current on the applied voltage bias. Just like any observable, the current also exhibits fluctuations around its average value, usually refereed to as {\it noise}. As this term suggests, noise was originally regarded to as a disturbing feature, possibly jeopardizing the stability of the signal in an electronic circuit. In recent years, however, especially in the field of mesoscopic physics, Rolf Landauer's famous remark that noise can be regarded as a signal too, has been substantiated by demonstrating that current noise contains remarkably useful information about the conductor. \cite{review1,review2,review3} For fractional quantum Hall systems, for instance, it has been shown that noise measurements allow to observe the  quasiparticle fractional charge characterizing the edge channels \cite{fract-charge1,fract-charge2}. More recently, in the field of nanoelectromechanical systems (NEMS)  it has been realized that, under appropriate conditions, noise can even be exploited to cool a mechanical resonator \cite{nems1,nems2}. \\

The  current noise at a measurement position $x$ and at a frequency $\omega$ is defined as
\begin{equation}
S^+(\omega,V;x)= \int_{-\infty}^{+\infty} \! \! \! dt \, e^{i \omega t} \,
\langle  \Delta \hat{J}(x,t)  \Delta \hat{J}(x,0)   \rangle    \label{S-non-sym-def}
\end{equation}
where $\Delta \hat{J}(x,t)=\hat{J}(x,t)-\langle \hat{J} \rangle$ is the current fluctuation operator, and $V$ is the bias voltage applied to the conductor. Eq.~(\ref{S-non-sym-def}) represents the non-symmetrized version of the noise, which  has recently attracted attention in mesoscopic
physics, for it can be directly measured by on-chip detectors 
\cite{Leso-Loos-JETP97,Aguado,gavish-imry}.\\

At equilibrium ($V=0$)  the average current through a conductor vanishes. Nevertheless, noise may be present because of thermal fluctuations (arising at finite temperature) and   quantum fluctuations (arising at finite frequency) \cite{devoret-review}.
While the role of thermal noise (also known as Johnson-Nyquist noise) has been investigated since long \cite{Johnson-Nyquist1,Johnson-Nyquist2}, the     regime $\hbar \omega > k_B T$, where quantum mechanics plays an important role,  has become accessible only more recently  thanks to advances in the design of high frequency electronics operating at low temperatures \cite{deblo03,Schoelkopf-science}.\\

We shall focus here on the purely quantum-mechanical  regime, and consider henceforth the case of zero temperature. Quantum fluctuations become manifest in a non-vanishing absorption noise spectrum of the conductor, i.e. $S^+(\omega,0;x) \neq 0$ for $\omega > 0 $, and originate from processes where  electrons absorb an energy quantum $\hbar \omega$ from the environment, typically a spectrum detector coupled to the conductor \cite{Buttiker-92}. When the system is driven out of equilibrium by  an applied voltage bias $V$,  an additional source of noise arises, due to the quantization of the electronic charge leading to a partitioning of electron scattering \cite{schottky,shot-noise1,shot-noise2,heiblum-reznikov-95}.\\

 On the experimental level, current and noise measurements are usually performed with an amplifier, which in turn exhibits its own intrinsic noise, often higher than the noise of the conductor to be measured. In order to avoid this problem and improve the signal-to-noise ratio,   it is favorable to measure the excess noise, {\it i.e.} the difference between the noise in presence of a finite bias voltage $V$ and the equilibrium noise
\begin{equation}
S_{\rm EX}(\omega,V;x) \doteq S^+(\omega,V;x)-  S^+(\omega,0;x)\, .
\end{equation}
The excess noise can also be expressed as
\begin{equation}
S_{\rm EX}(\omega,V;x)= \int_0^V \frac{\partial S^+(\omega,V^\prime;x)}{\partial V^\prime} dV^\prime \label{SEX-as-int}\, .
\end{equation}
In view of the additional noise generating processes arising in presence of a finite voltage $V$, excess noise is usually expected to be a positive quantity. This intuition is, however, wrong in general. Indeed Lesovik and Loosen \cite{Leso-Loos} have  pointed out that in the regime $eV \gg k_B T$ the shot noise $S^{+}(0,V;x)$ becomes smaller than the equilibrium Johnson-Nyquist noise when  the  transmission coefficient of a conductor is
non-vanishing only in an energy window $\delta E$, the Fermi energy lies within this energy window, and the temperature is sufficiently high ($k_B T \gg \delta E$). More recently, it has been shown that the excess noise can be negative also in the quantum regime \cite{sex-paper}, implying the formation of an out of equilibrium state with noise smaller than the equilibrium  quantum noise. \\

To clarify the conditions leading to negativity of the excess noise, it is natural to inquire whether the noise measured at a given applied voltage bias $V$ can be tuned. One possibility to tune the noise is to change the number of conducting channels by applying a gate voltage to the conductor, as is routinely done, for instance, with quantum point contacts \cite{heiblum-reznikov-95}. This requires a multi-channel conductor and induces nearly discrete changes. In contrast, here we look for lines in the parameter space separating regimes of different sign of the excess noise, so that continuous tuning would be desirable.
To this purpose, a suitable system  is a quantum  interferometer. Indeed,   interferometers  offer
the appealing feature   that   scattering properties can be tuned via the interference conditions, opening the possibility to modify the current noise while keeping  the voltage $V$ (and possibly also the frequency $\omega$) fixed. In this paper, using a concrete model system, we shall show that excess noise can indeed be tuned to negative values.
\begin{figure}
\centering
\includegraphics[width=0.70\linewidth]{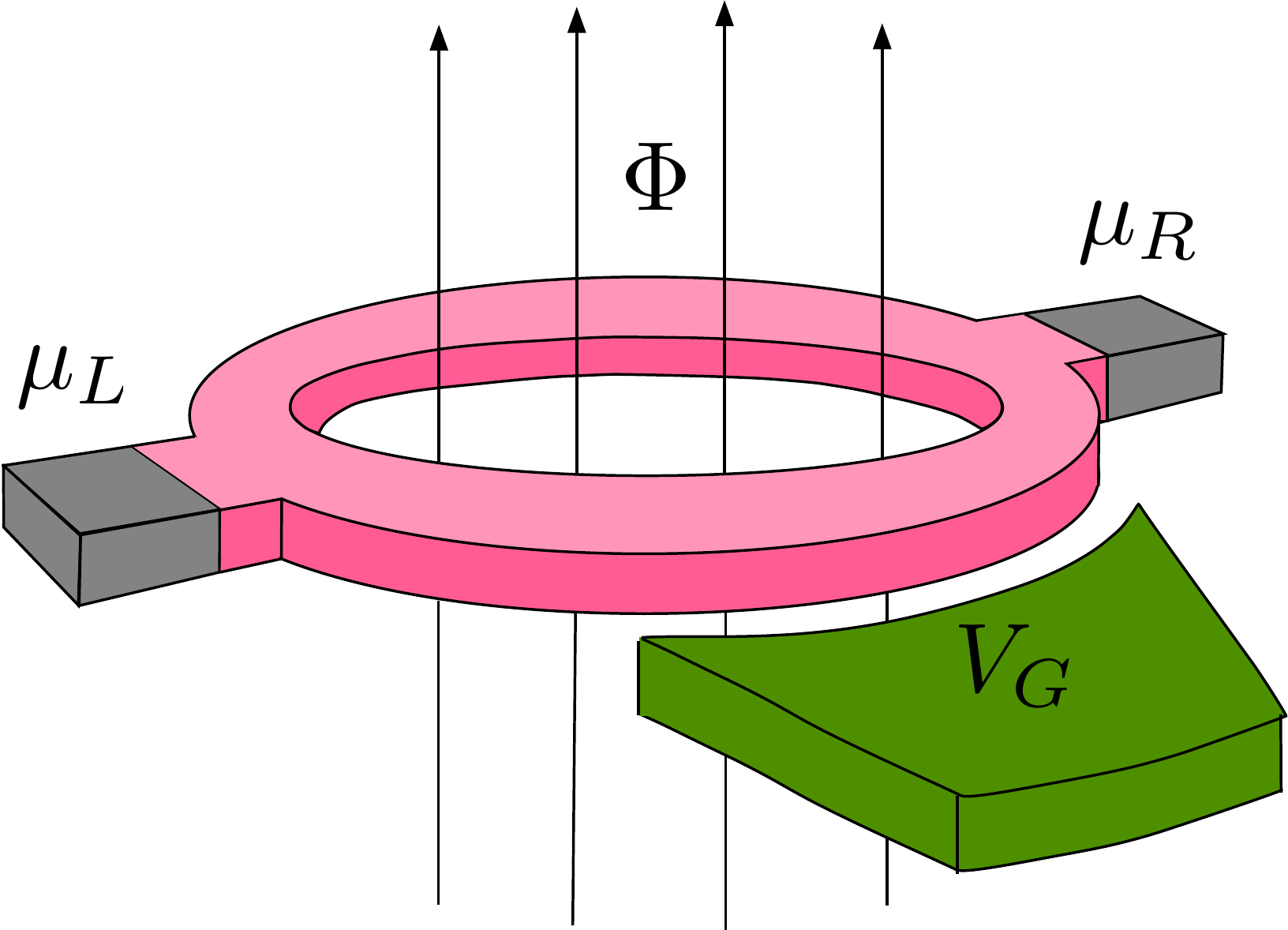}
\caption{Sketch of the Aharonov-Bohm interferometer. A ring threaded by a magnetic flux $\Phi$ is connected to two metallic electrodes at electrochemical potentials $\mu_L$ and $\mu_R$ implying a voltage bias $V=(\mu_L-\mu_R)/e$. A side gate with gate voltage $V_G$ is coupled to one of the two arms of the interferometer.}
\end{figure}

\section{The model}
\begin{figure}
\centering
\includegraphics[width=\linewidth]{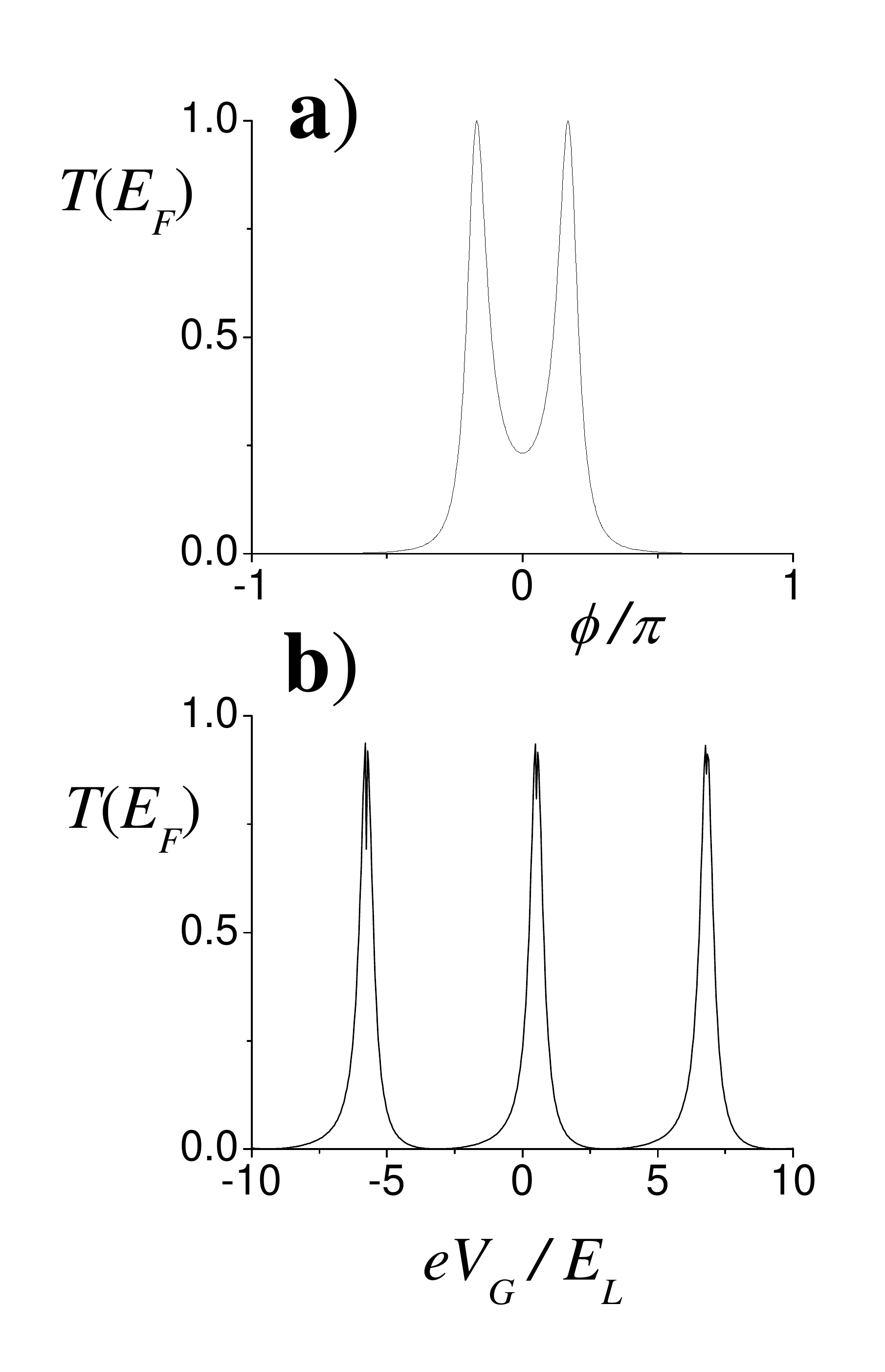}
\caption{The transmission coefficient of the Aharonov-Bohm ring at the Fermi energy $E_F = 100\, E_L$ is plotted as a function of (a) the dimensionless magnetic flux $\phi$ and (b) the side gate voltage $u=eV_G/E_L$.}
\end{figure}
In order to illustrate the effect of noise tuning, we shall consider an Aharonov-Bohm interferometer, consisting of a ring contacted by two electrodes, as depicted in Fig. 1. The ring is threaded by a magnetic flux $\Phi$; furthermore we envisage a side gate which changes the potential of the (say) lower arm with respect to the upper arm by an amount $V_G$. The interference condition between electron propagation in the two arms can therefore be controlled both magnetically, through the flux, and electrostatically, through the side gate voltage $V_G$. For simplicity we assume that the ring is symmetric and narrow so that the electrons propagate in a single channel. The contacts of the ring to the electrodes are modeled as Y-junctions, characterized by the customary scattering matrix \cite{buttiker-ring-matrix1,buttiker-ring-matrix2}

\begin{equation}
\mathsf{S}_Y(\eta)=
\left( \begin{array}{ccc}
\displaystyle -\sqrt{1-2 \eta} & \sqrt{\eta} & \sqrt{\eta} \\
 \sqrt{\eta} & a(\eta) & b(\eta) \\
 \sqrt{\eta} & b(\eta) & a(\eta)   \end{array} \right)\label{SY}
\end{equation}
with
$a(\eta) =   (\sqrt{1-2\eta}-1)/2$ and $b(\eta)=(\sqrt{1-2\eta}+1)/2$.
Here $\eta \in [0; 1/2] $ is a parameter interpolating between total reflection and perfect transmission
of an electron approaching the Y-junction from the lead. Propagation of the electron wave function along the two arms of the ring involves phase factors that depend on the magnetic flux $\Phi$, as well as on the side gate voltage $V_G$ applied to the lower arm. For simplicity, we shall neglect here flux and side gate voltage fluctuations, which have been addressed elsewhere~\cite{flux-fluct1,flux-fluct2,flux-fluct3}.\\

Combining the  scattering matrices (\ref{SY}) at the two contacts with the propagation along the arms, a rather lengthy but straightforward calculation yields the scattering matrix for the Aharonov-Bohm interferometer
\begin{equation}
\mathsf{S}_E=\left(
\begin{array}{cc}
 r^{}_E & t^\prime_E \\
 t^{}_E & r^\prime_E
\end{array} \right)  \label{S-matr}
\end{equation}
where $E$ denotes the energy of the incident electron and $r$ and $r^\prime$ ($t$ and $t^\prime$) are the reflection (transmission) amplitudes.
These matrix elements can be suitably expressed as~\cite{dolcini-giazotto}  $r=r^\prime=\rho_P/P^*$, $t=-i \tau^*_P/P^*$, and $t^\prime=-i \tau_P/P^*$, with
\begin{eqnarray}
P(\varepsilon)&=& \left[ e^{i (\varepsilon+\kappa_F +\frac{u}{2})} -  e^{-i (\varepsilon+\kappa_F +\frac{u}{2})} \cos{\gamma}\right]^2   \nonumber \\
& & +\, 4 \cos{\gamma} \sin^2\left(\frac{{\phi}}{2}\right) \label{P-def} \\
& & -\,(1-\cos{\gamma})^2 \prod_{s=\pm}\cos \left( \frac{u+ s {\phi}}{2} \right)\, , \nonumber
\end{eqnarray}
\begin{eqnarray}
\tau_{P}(\varepsilon)  &=& -i\sin^2{\gamma}  \prod_{s=\pm} s \,  e^{i \, s \, (\varepsilon+\kappa_F +\frac{u}{2})}
\nonumber\\ &&\times\ \cos \left( \frac{u+s {\phi}}{2} \right)  \, ,
\end{eqnarray}
and
\begin{eqnarray}
\rho_{P}(\varepsilon)&=& 4 \cos{\gamma} \, \prod_{s=\pm} \sin\left(\varepsilon+\kappa_F +\frac{u+s{\phi}}{2}\right)  \nonumber \\
& & \, + \, (1-\cos{\gamma})^2 \prod_{s=\pm}\sin \left( \frac{u+ s {\phi}}{2} \right)\, .
\end{eqnarray}
Here $\varepsilon=(E-E_F)/E_L$ describes the deviation  of the energy $E$ from the Fermi level $E_F$,  expressed in terms of the ring ballistic frequency
\begin{equation}\label{EL}
    E_L = \hbar v_F/L\, ,
\end{equation}
where $L$ is the length of either ring arm; $u=eV_G/E_L$ is the dimensionless side gate voltage,   $\kappa_F =k_F L$ is the dimensionless Fermi wavevector, and $\phi=2\pi \Phi/\Phi_0$ is the dimensionless flux, with $\Phi_0=h/e$ denoting the   flux quantum.
Finally, $\gamma \in [0;\pi/2]$ is a convenient re-parametrization of the Y-junction transparency
$T_Y=\sin^2{\gamma}$ which is related with $\eta$ in Eq.~(\ref{SY}) through the relation $\eta=(\sin^2 \gamma)/2$.
The spectrum of the ring has been linearized around the equilibrium Fermi energy $E_F$, under the realistic assumption  that all relevant electronic energies deviate from the Fermi energy by an energy difference much smaller than $E_F$. We note in passing that the presence of the magnetic flux $\phi$ makes the scattering matrix (\ref{S-matr}) non-symmetric due to the breaking of time-reversal symmetry. \\

The tunability of the Aharonov-Bohm interferometer is explicitly shown in Fig.~2, where the  transmission coefficient $T_E=|t_E|^2=|t_E^\prime|^2$ of the ring at the Fermi energy $E_F$ is shown as a function of the magnetic flux [see Fig. 2(a)] and the side gate voltage [see Fig. 2(b)].
The current operator at a measurement point located  in the (say) left lead reads
\begin{eqnarray}
\hat{J}(x,t) &=& \frac{e}{2 \pi \hbar } \int\limits_{-\infty}^{\infty} \int\limits_{-\infty}^{\infty} dE \, dE'   e^{i (E-E') t/
\hbar}  \label{JL-op} \\
& & \times\! \sum_{X,Y=L/R} \left\{ A_{L}^{X Y}(E,E';x)  :{\hat{a}}^{\dagger}_{X E}\hat{a}^{}_{Y E'}:   \right\} \nonumber
\end{eqnarray}
where the symbol $: \, \, \, :$ denotes normal ordering with respect to the equilibrium Fermi sea, the label $L$ [$R$] refers to the left [right] lead, and the dimensionless coefficients $A_L^{X Y}$ are related to the elements of the scattering matrix~(\ref{S-matr}) by
\begin{equation}
 \begin{array}{lcl}
\displaystyle A_L^{LL}(E,E^\prime;x) &=& e^{-i (k_E-k_{E^\prime}) x} - e^{i (k_E-k_{E^\prime}) x} r^{*}_E  r^{}_{E^\prime} \\ & & \\
\displaystyle A_L^{RR}(E,E^\prime;x) &=&  - e^{i (k_E-k_{E^\prime}) x} {t^\prime}^*_{E} t^\prime_{E^\prime} \\ & & \\
\displaystyle A_L^{LR}(E,E^\prime;x) &=& - e^{i (k_E-k_{E^\prime}) x} r^{*}_E t^\prime_{E^\prime}\,  \\ & & \\
\displaystyle A_L^{RL}(E,E^\prime;x) &=& - e^{i (k_E-k_{E^\prime}) x} {t^\prime}^*_{E} r^{}_{E^\prime}\,.
\end{array} \label{A_L-def}
\end{equation}
Substituting this expression into the definition (\ref{S-non-sym-def}) of the non-symmetrized current noise at a measurement point $x$ located in the (say) left electrode, one obtains
\begin{eqnarray}
\lefteqn{ S^{+}(\omega,V;x) = \frac{e^2}{2 \pi \hbar}
\int\limits_{-\infty}^{\infty}  dE  \sum_{X,Y=L,R} } & & \label{S+LL} \\
 & &    |A_{L}^{XY}(E,E+\hbar
\omega;x) |^2 f_X(E) (1-f_Y(E+\hbar \omega))\, ,
\nonumber
\end{eqnarray}
where $f_{L/R}(E)=f(E-\mu_{L/R})$ denotes the Fermi distribution function in the $L/R$  lead with the electrochemical potential
$
\mu_{L/R}=E_F+ e \, a_{L/R} V $. Here
\begin{equation}
V = (\mu_L - \mu_R)/e
\end{equation}
is the applied voltage, while $a_L$ and  $a_R=a_L-1$ are  numerical coefficients depending on the biasing scheme. For symmetric bias $a_L=a_R=1/2$ while for completely asymmetric bias one of the coefficients vanishes. \\
A straightforward calculation also yields the voltage derivative of the noise at zero temperature
\begin{eqnarray}
\lefteqn{ \frac{\partial S^{+}}{\partial V} (\omega,V;x) = \frac{e^3}{2
\pi \hbar}  \, \cdot }  \label{dSdVuno} \\
& & \bigg\{ \theta(\omega)  \sum_{X=L,R}  a_X\sum_{s=\pm 1} s  \,
 |A_{L}^{XX}(\mu_X,\mu_X+s\hbar \omega;x)|^2 \,
  \nonumber \\
& & \ \ + \theta(\hbar \omega+eV)  \Big[ a_L \,
  |A_{L}^{LR}(\mu_L,\mu_L+\hbar \omega;x)|^2     \nonumber  \\
& & \ \ \hspace{1.9cm}  -a_R \, |A_{L}^{LR}(\mu_R-\hbar
\omega;\mu_R;x)|^2 \,
  \Big]    \nonumber \\ & & \nonumber \\
& &\ \  - \theta(\hbar \omega-eV)  \Big[  a_L
 \,    |A_{L}^{RL}(\mu_L-\hbar \omega;\mu_L;x)|^2 \,
  \nonumber \\
& &\ \  \hspace{1.9cm}  - a_R \,
|A_{L}^{RL}(\mu_R,\mu_R+\hbar \omega;x)|^2 \,    \Big] \bigg\}
\nonumber
\end{eqnarray}
and the excess noise is easily computed from Eq.~(\ref{SEX-as-int}).\\

As one can see, the noise depends on the  coefficients $A_L^{XY}$ which are related to the scattering matrix of the mesoscopic system [see Eq.~(\ref{A_L-def})].   In particular,  we shall show below that, because of competing signs in front of the various terms in Eq.~(\ref{dSdVuno}), the excess noise may be driven to negative values.
\section{Results and Discussion}
\subsection{Shot noise}
We start our analysis with the case $\omega=0$ (shot noise): at zero frequency (and zero temperature) the equilibrium noise vanishes and the excess noise coincides with the  shot noise. Now, from Eqs.~(\ref{A_L-def}) and (\ref{S+LL}) we see that the shot noise
\begin{equation}\label{S-a-tOmega-Zeor}
    S^{+}_0(V)\equiv S^{+}(\omega=0,V;x)
\end{equation}
is independent of the measurement position $x$ since the same is true for the coefficients $A_L^{XY}(E,E^{\prime};x)$ for $E=E^{\prime}$.
The Fano factor
\begin{equation}
F=\frac{S^+_0(V)}{e I}
\end{equation}
quantifies the ratio of the noise to the average current. To illustrate the benefit of the Aharonov-Bohm interferometer, we first focus on the regime of small applied bias $eV \ll E_L \ll E_F$: In this case the transmission and reflection coefficients depend only very weakly on energy, and one recovers the well known  expressions $I=(e^2/h) \, T V$ for the average current and $S^+_0= (e^3/h) T (1-T) V$ for the shot noise. Exploiting the dependence of the transmission coefficient $T$ on the magnetic flux and the side gate potential, it is thus possible to tune the  transport conditions,   for example, to a minimum of the Fano factor, as shown in  Fig.~3. Similar results have also been  obtained within a tight-binding approach~\cite{davidovich}, and for the case of  quantum dots  embedded in the arms of an Aharonov-Bohm interferometer~\cite{AB-dots1,AB-dots2}.\\
\begin{figure}
\centering
\includegraphics[width=0.9\linewidth]{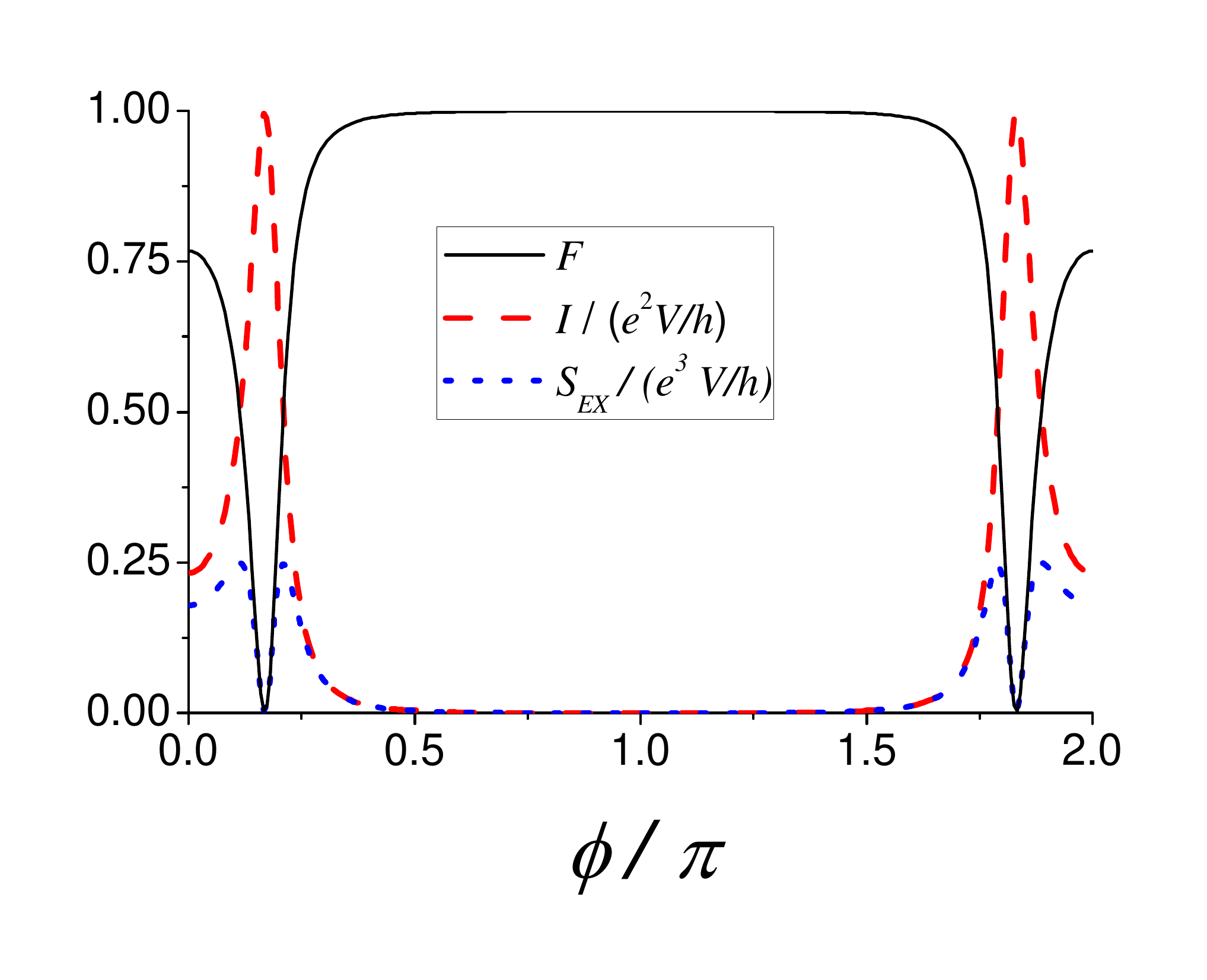}
\caption{The Fano factor (solid curve), the dimensionless current (dashed curve) and the shot noise (dotted curve) of an Aharonov-Bohm interferometer biased with a voltage $eV=E_L/100$ are plotted as a function of the dimensionless magnetic flux. The ring parameters are $\gamma=\pi/6$, $E_F=100\, E_L$, and $a_L=1/2$.}
\end{figure}

In general, if the applied voltage is comparable with (or bigger than) the ballistic ring energy $E_L$, the energy dependence must be taken into account and the noise is readily obtained from Eq.~(\ref{SEX-as-int}) upon integration of  the expression
\begin{eqnarray}
\lefteqn{ \frac{\partial
S^{+}}{\partial V}(\omega=0,V;x)=   \frac{e^3}{h} \,
\mbox{sgn}(V)   } & & \label{dSdV-shot} \\
& & \times \left[ a_L \,
  T(\mu_L) R(\mu_L)     \, - \, a_R \, T(\mu_R) R(\mu_R)
  \right]
\nonumber
\end{eqnarray}
which follows from Eq.~(\ref{dSdVuno}) for $\omega=0$.
\subsection{Finite frequency noise}
Let us now consider the case of finite frequency $\omega \neq 0$. Then the equilibrium noise $S^+(\omega;0;x)$ is non-vanishing due to quantum fluctuations, so that the excess noise $S_{\rm EX}$ differs  from the total current noise even at zero temperature.
Crucial differences with respect to the zero frequency case emerge  mainly in view of two important aspects. In the first instance, the energy dependence of the scattering matrix may become important even in the  regime of very small voltage bias, if the frequency is comparable with the ring ballistic frequency $E_L/\hbar$. Secondly,   finite frequency   measurements are able to resolve the internal dynamics of the conductor  and are thus sensitive to instantaneous charge fluctuations caused by current flow. These fluctuations induce instantaneous changes of the potential profile $U$ inside the conductor, which in turn affect the  scattering properties. Because of this feed-back process, the determination of finite
frequency transport properties is essentially an {\it interacting} problem \cite{Buttiker-screening1,Buttiker-screening2,Buttiker-screening3} for which the potential profile $U$ must be determined self-consistently.  \\

\begin{figure}
\centering
\includegraphics[width=0.9\linewidth]{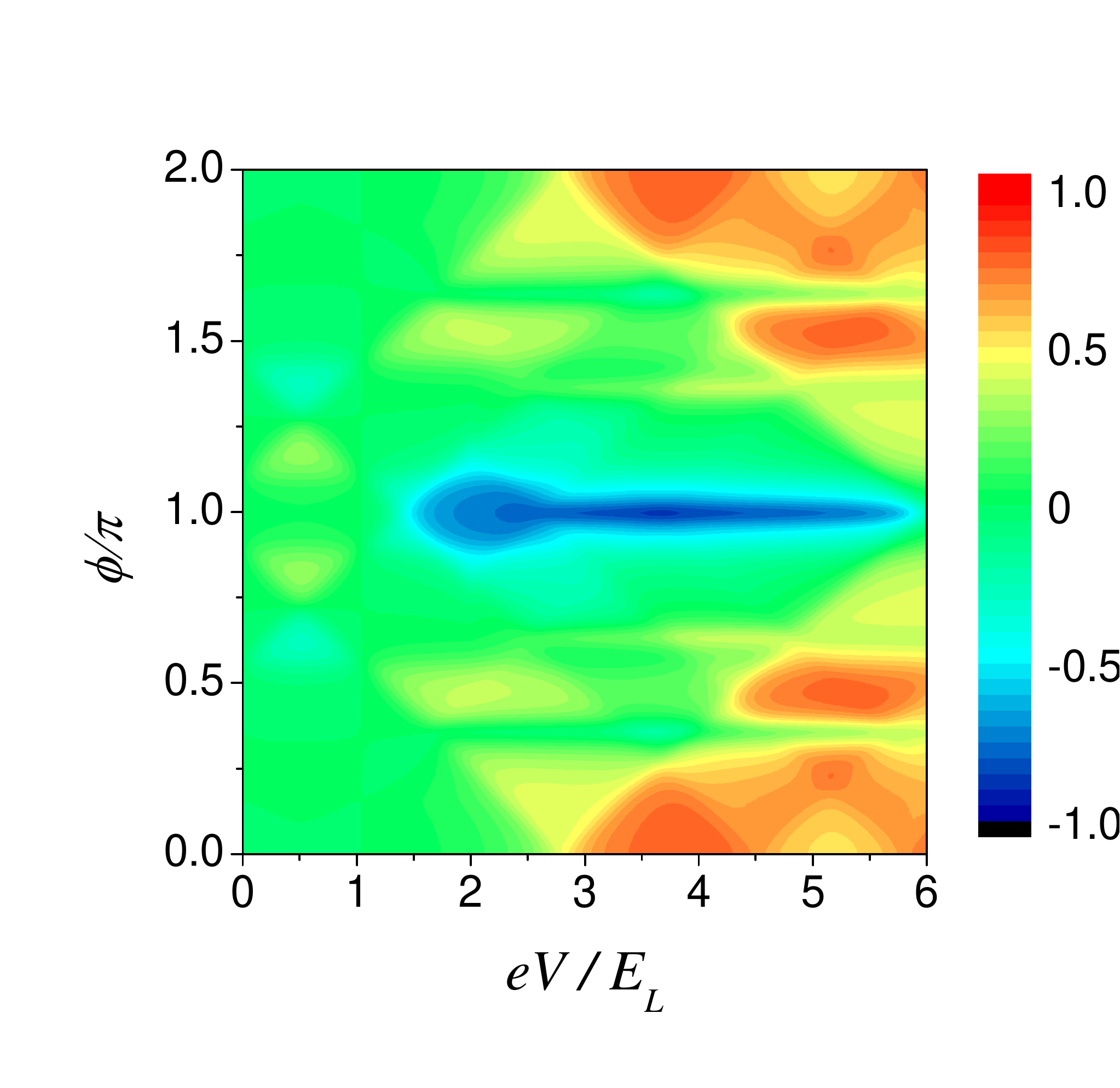}
\caption{The excess noise (in units of $e^2 E_L/h$) at finite frequency $\omega=2 E_L/\hbar$ is plotted as a function of the applied bias voltage $V$ and the dimensionless magnetic flux $\phi =2 \pi \Phi/\Phi_0$.  Regions of negative excess noise can be reached by magnetic tuning of the Aharonov-Bohm ring flux.  The ring parameters are $\gamma=\pi/6$, $E_F=100 E_L$, $a_L=1/2$, $V_G=0$, and the measurement point is at the left contact.}
\end{figure}
\begin{figure}
\centering
\includegraphics[width=0.9\linewidth]{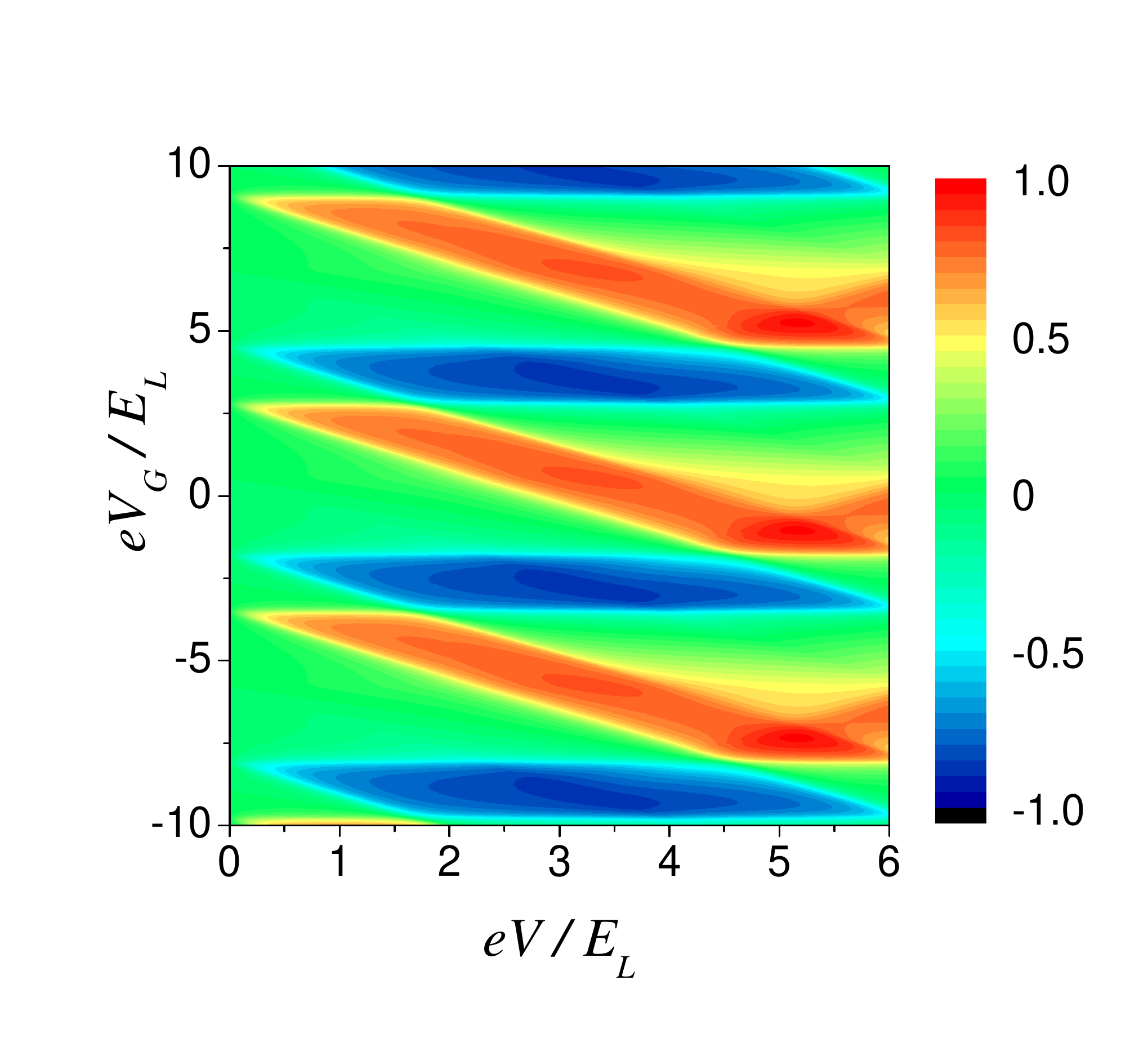}
\caption{The excess noise (in units of $e^2 E_L/h$) at finite frequency $\omega=2 E_L/\hbar$ is plotted as a function of the applied bias voltage $V$ and the side gate voltage  $V_G$.   Regions of negative excess noise can be reached by electrostatic tuning of the side gate potential. The ring parameters are $\gamma=\pi/6$, $E_F=100 E_L$, $a_L=1/2$, $\phi=0$, and the measurement point is at the left contact.}
\end{figure}

The above-mentioned factors affect the finite frequency noise in unlike ways, thus leading to rather rich finite frequency current noise features. A specific question we wish to address here is whether the out of equilibrium noise may become smaller than the equilibrium noise, thus leading to negative excess noise.
The investigation of this problem is quite difficult in general, because various factors interplay, in particular, the energy dependence of the scattering matrix, the frequency regime under investigation, and  the internal screening properties of the conductor. Some approximations are thus necessary in order to proceed. A simple but effective method to account for the instantaneous potential profile fluctuations $\Delta U$ in terms of a single parameter is based on the geometrical capacitance $C$  relating the  spectrum $\Delta U(\omega)$ of potential variations  to the  electronic charge spectrum $Q(\omega)$ through the relation $\Delta U(\omega) \sim Q(\omega)/C$. The importance of potential profile fluctuations thus strongly   depends on the size of the geometrical capacitance $C$, which has to be compared with the intrinsic quantum capacitance $C_q$ of the conductor, related to its density of states.

In a recent paper \cite{sex-paper} investigating the excess noise of a single channel quantum wire, we  have shown that the excess noise  becomes negative when the geometrical capacitance is smaller than 1/3 of the quantum capacitance. Here, instead, we shall focus on the case of a large geometrical capacitance $C \gg C_q$, when fluctuations of the charge of the ring induce negligible variations in the potential $U$, so that the independent electron approach remains valid. Nevertheless, since transport is analyzed at finite frequency, the energy dependence of the scattering matrix is important\cite{ines-arXiv}. Indeed, the noise   $S^+(\omega,V;x)$, which is always positive,  can be expressed as an integral of  scattering matrix elements $|A_L^{XY}|^2$ over energy windows determined by the Fermi functions of the leads [see Eq.(\ref{S+LL})]. Importantly, the out of equilibrium noise and the equilibrium quantum noise are characterized by different energy windows, because the former depend  on the applied voltage~$V$, besides the finite frequency~$\omega$. By varying the energy profile of the scattering matrix via Aharonov-Bohm interferometry,   the out of equilibrium noise may thus become larger or smaller than the noise at $V=0$. Concrete result are shown   in Fig.~4,  where the finite frequency excess noise is plotted as a function of the applied voltage bias~$V$ and the magnetic flux $\phi$. Regions of negative excess noise are present, and can be reached simply by varying   $\phi$. Fig.~5 describes the same effect, yet, as a function of the side gate voltage $V_G$. These results show that in an Aharonov-Bohm interferometer the sign of the excess noise can be tuned either magnetically or electrostatically.\\

To conclude, we wish to shortly discuss   the order of magnitude of the parameters involved. Experimental evidence for Aharonov-Bohm oscillations has been found in semiconductor-based rings in the mesoscopic regime~\cite{ring1,ring2,ring3,ring4,ring5,ring6} and, quite recently, in graphene-based rings~\cite{graphene-rings} in the mesoscopic regime. With respect to metallic rings, semiconducting structures offer the advantage of exhibiting a dependence on the side gate voltage, thus being electrostatically tunable in the same way as discussed in our model. The typical size $L$ of such rings is about $1 {\mu \rm m}$, which corresponds to an energy $E_L$ of about 1 meV, or equivalently to  a ballistic frequency $E_L/\hbar$ of about 1 THz. Although this frequency regime was far from   experimental reach  two decades ago, more recent high frequency measurements~\cite{Aguado,deblo03} show that nowadays the THz range is no longer  unrealistic. It thus seems likely that our proposal for tuning the sign of the excess noise can be tested in the near future.

\section{Acknowledgements}
One of the authors (H.G.) wishes to thank Peter H{\"a}nggi for a stimulating cooperation during the days when we both moved as young researchers between Stuttgart, Basel, Philadelphia, New York, San Diego, Rehovot, and elsewhere. The authors acknowledge support by the Excellence Initiative of the German Federal and State Governments.

\end{document}